\documentclass[page-classic]{epl2} 
\usepackage{amssymb}

\usepackage{graphicx}
\usepackage{dcolumn}
\usepackage{bm}

\title{Solutions to the quantum Rabi model with two equivalent qubits}
\date{\today}
\author{Hui Wang$^{1}$, Shu He$^{1,2}$, Liwei Duan$^{1}$, Yang Zhao $^{3}$,
and Qing-Hu Chen$^{1,2,*}$}

\institute{
\inst{1} Department of Physics, Zhejiang University, Hangzhou 310027,
P. R. China\\
\inst{2} Center for Statistical and Theoretical Condensed Matter
Physics, Zhejiang Normal University, Jinhua 321004, P. R. China\\
\inst{3} Division of Materials Science, Nanyang Technological
University, 50 Nanyang Avenue, Singapore 639798
 }


\date{\today}
\abstract{ Using extended coherent states, an analytically exact
study has been carried out for the quantum Rabi model with two
equivalent qubits. Compact transcendental functions of one variable
have been derived leading to exact solutions. The energy spectrum is
clearly identified and analyzed. Also obtained analytically is the
necessary and sufficient conditions for the occurrence of isolated
exceptional solutions, which
are not doubly degenerate as in the one-qubit quantum Rabi model. }
\begin{document}

\maketitle

\section{Introduction}

Quantum Rabi model (QRM) describes a two-level atom (qubit) coupled to a
cavity electromagnetic mode (an oscillator)\cite{Rabi}, a minimalist
paradigm of matter-light interactions with applications in numerous fields
ranging from quantum optics, quantum information science to condensed matter
physics. The solutions to the QRM are however highly nontrivial. Recently,
Braak presented an analytically exact solution \cite{Braak} to a one-photon
QRM using the representation of bosonic creation and annihilation operators
in the Bargmann space of analytical functions \cite{Bargmann}. A so-called G
function with a single energy variable was derived yielding exact
eigensolutions. Alternatively, using the method of extended coherent states
(ECS), this G function was recovered in a simpler, yet physically more
transparent manner by Chen \textit{et al.}\cite{Chen2012}.

The extension of QRM to multiple qubits coupled with a single cavity mode,
also known as the Dicke model\cite{Dicke}, has also attracted considerable
attention both theoretically \cite
{Emary,liberti,vidal,chen1,Bastarrachea,Agarwal} and experimentally \cite
{Scheibner,Schneble,Hartmann,Nataf,Vukics,Baumann} in the past decade. As
quantum information resources such as the quantum entanglement \cite{Nielsen}
and the quantum discord~\cite{ollivier} can be easily stored in two qubits
and the Greenberger-Horne-Zeilinger (GHZ) states \cite{Greenberger} which
are generated in three qubits, it is not suprising that devices with more
than one qubit find potential applications in quantum information technology%
\cite{Sillanpaa}. More recently, some exact analytical solutions are
attempted for a small number of qubits, such as two qubits\cite
{Chilingaryan,Rodriguez,Peng} and three equivalent qubits\cite{Braak1} in
the representation of the Bargmann space. In those studies, the G functions
are often determined by matrices of high dimensions, and are therefore more
complicated in form than that in the one-qubit QRM \cite{Braak}. It is also
a tedious task to get eigensolutions by solving coupled, highly nonlinear
equations of multiple variables\cite{Chilingaryan,Rodriguez,Peng,Trav}. In
the absence of a simple G function, it is difficult to arrive at a concise
description of the spectrum, and all but impossible to derive analytically a
condition for the occurrence of exceptional solutions\cite{Trav}, which
exist at special values of model parameters, and are the eigenvalues that do
not correspond to zeros of the G function.

In this work, employing ECS, we demonstrate a successful derivation of a
concise G function for the QRM with two equivalent qubits, similar to the
case of the one-qubit QRM \cite{Braak,Chen2012}. Moreover, some isolated
exact solution and the necessary and sufficient condition for its
occurrence, similar to the Juddian solutions \cite{Judd}, are also presented.

The remainder of the paper is organized as follows. In Sec. II, a single
variable G function is derived in detail for the QRM with two equivalent
qubits. The solutions of these G functions and discussions about the
exceptional solutions are presented in Sec. III, a brief summary is given
finally.

\section{The G-function}

The Hamiltonian of the QRM with two equivalent qubits, also known as the
Dicke model of $N=2$, can be written as
\begin{equation}
H={\omega }d^{\dag }d+\Delta J_{z}+\frac{2\lambda }{\sqrt{N}}(d^{\dag
}+d)J_{x},  \label{Hamiltonian}
\end{equation}
where $\Delta $ is the energy splitting of the two-level system (qubit), $%
d^{\dag }$ creates one photon in the common single-mode cavity with
frequency $\omega $, $\lambda $ describes the atom-cavity coupling strength,
and
\begin{equation}
J_{z}=\frac{1}{2}\sum_{i=1,2}\sigma _{z}^{(i)},J_{x}=\frac{1}{2}%
\sum_{i=1,2}\sigma _{x}^{(i)}=\frac{1}{2}\left( J_{+}+J_{-}\right),
\end{equation}
where$\ $ $\sigma _{x,z}^{(i)}~(i=1,2)$ is the Pauli operator of the $i-$th
qubit, $J$ is the usual angular momentum operator, and $J_{+}$ and $J_{-}$
are the angular rising and lowering operators and obey the $SU(2)$ Lie
algebra $[J_{+},J_{-}]=2J_{z},[J_{z},J_{\pm }]=\pm J_{\pm }$. So the Hilbert
space of this algebra in this model is spanned by the Dicke state $%
\left\vert j,m_{z}\right\rangle $ with $j=1,m_{z}=-1,0,1$ and $j=0,m_{z}=0$.

There is a trivial exact eigensolution in the subspace of the $j=0$ Dicke
state, because $\left\vert j=0,m_{z}=0\right\rangle $ is simultaneously the
eigenstate of $J_{z}\ $and $J_{x}$. The eigenfunction of the whole system in
this subspace is $\left\vert n\right\rangle =\left\vert 0,0\right\rangle
\left\vert n\right\rangle _{\mathrm{ph}}$ where $\left\vert n\right\rangle _{%
\mathrm{ph}} =\frac{\left( d^{\dagger }\right) ^{n}}{\sqrt{n!}} \left\vert
0\right\rangle $ with eigenvalue $E_{n}=n\hbar \omega $. The solution in the
space of the $j=1$ eigenstate is however highly nontrivial, and has been a
subject of recent interest\cite{Agarwal,Chilingaryan,Peng}. Our goal is to
find analytically exact solutions with simplest forms.

A transformed Hamiltonian with a rotation with respect to the $y$ axis by an
angle $\frac{\pi }{2}$ in the matrix form can be written as ( in units of $%
\hbar =\omega =1$)
\begin{equation}
H=\left(
\begin{array}{ccc}
d^{\dag }d+g(d^{\dag }+d) & -\frac{\Delta }{\sqrt{2}} & 0 \\
-\frac{\Delta }{\sqrt{2}} & d^{\dag }d & -\frac{\Delta }{\sqrt{2}} \\
0 & -\frac{\Delta }{\sqrt{2}} & d^{\dag }d-g(d^{\dag }+d)
\end{array}
\right),
\end{equation}
where $g=\frac{2\lambda }{\sqrt{2}}$.

In this paper, we analytically study the $N=2$ Dicke model with similar
ansatz of the wavefunctions as in Ref. \cite{chen1}. Note that the diagonal
elements in the above matrix can be changed into free particle number
operators by shifting the photonic operator with displacements $g,0,-g $. To
employ our previous ECS approach\cite{chen94,chen1}, we perform the
following two Bogoliubov transformations with finite displacement
\begin{equation}
A=d+g,B=d-g.
\end{equation}
In Bogoliubov operators $A$($B$), the matrix element $H_{11}$ ($H_{33}$) can
be reduced to the free particle number operators $A^{\dagger }A$ ($%
B^{\dagger }B$) plus a constant, which is very helpful for the further study.

First, the wavefunction is proposed in terms of operator $A$ as
\begin{equation}
\left| {}\right\rangle =\left(
\begin{array}{c}
\sum_{n=0}^\infty \sqrt{n!}u_n|n\rangle _A \\
\sum_{n=0}^\infty \sqrt{n!}v_n|n\rangle _A \\
\sum_{n=0}^\infty \sqrt{n!}w_n|n\rangle _A
\end{array}
,\right)  \label{wave_A}
\end{equation}
where $u_n,v_n,$and $f_n$ are the expansion coefficients, $\left|
n\right\rangle _A\;$is just called extended coherent state with the
following properties
\begin{eqnarray}
\left| n\right\rangle _A &=&\frac{\left( A^{\dagger }\right) ^n}{\sqrt{n!}}%
\left| 0\right\rangle _A=\frac{\left( d^{\dagger }+g\right) ^n}{\sqrt{n!}}%
\left| 0\right\rangle _A, \\
\left| 0\right\rangle _A &=&e^{-\frac 12g^2-gd^{\dagger }}\left|
0\right\rangle .
\end{eqnarray}
where the vacuum state $\left| 0\right\rangle _A$ of the Bogoliubov
operators $A$ is a well-defined eigenstate of the one-photon annihilation
operator $d$, also known as a coherent state or a displaced oscillator \cite
{Glauber}.

The Schr\"{o}dinger equation leads to
\begin{eqnarray*}
E\sum_{n=0}^\infty \sqrt{n!}u_n|n\rangle _A &=&\sum_{n=0}^\infty \sqrt{n!}[%
(n-g^2)u_n-\frac \Delta 2\sqrt{2}v_n]|n\rangle _A, \\
E\sum_{n=0}^\infty \sqrt{n!}v_n|n\rangle _A &=&\sum_{n=0}^\infty \sqrt{n!}%
\{[(n+g^2)v_n-\frac \Delta 2\sqrt{2}(u_n+w_n)]|n\rangle _A \\
&&-g(\sqrt{n+1}v_n|n+1\rangle _A+\sqrt{n}v_n|n-1\rangle _A)\}, \\
E\sum_{n=0}^\infty \sqrt{n!}w_n|n\rangle _A &=&\sum_{n=0}^\infty \sqrt{n!}%
\{[(n+3g^2)w_n-\frac \Delta 2\sqrt{2}v_n]|n\rangle _A \\
&&-2g(\sqrt{n+1}w_n|n+1\rangle _A+\sqrt{n}w_n|n-1\rangle _A)\}.
\end{eqnarray*}
Projecting the above three equations onto $_A\langle m|$ yields
\begin{eqnarray}
Eu_m &=&(m-g^2)u_m-\frac \Delta {\sqrt{2}}v_m, \\
Ev_m &=&(m+g^2)v_m-\frac \Delta {\sqrt{2}}(u_m+w_m)-gv_{m-1}-g(m+1)v_{m+1},
\\
Ew_m &=&(m+3g^2)w_m-\frac \Delta {\sqrt{2}}v_m-2gw_{m-1}-2g(m+1)w_{m+1}.
\label{coeff_A}
\end{eqnarray}
Similarly, using the second operator $B$, the wavefunction can be expressed
as
\begin{equation}
|\rangle =\left(
\begin{array}{c}
\sum_{n=0}^\infty (-1)^n\sqrt{n!}w_n^{\prime }|n\rangle _B \\
\sum_{n=0}^\infty (-1)^n\sqrt{n!}v_n^{\prime }|n\rangle _B \\
\sum_{n=0}^\infty (-1)^n\sqrt{n!}u_n^{\prime }|n\rangle _B
\end{array}
\right) ,  \label{wave_B}
\end{equation}
Proceeding as before, we have
\begin{eqnarray}
Eu_m^{\prime } &=&(m-g^2)u_m^{\prime }-\frac \Delta {\sqrt{2}}v_m^{\prime },
\\
Ev_m^{\prime } &=&(m+g^2)v_m^{\prime }-\frac \Delta {\sqrt{2}}(u_m^{\prime
}+w_m)^{\prime }-gv_{m-1}^{\prime }-g(m+1)v_{m+1}^{\prime }, \\
Ew_m^{\prime } &=&(m+3g^2)w_m^{\prime }-\frac \Delta {\sqrt{2}}v_m^{\prime
}-2gw_{m-1}^{\prime }-2g(m+1)w_{m+1}^{\prime }.  \label{coeff_B}
\end{eqnarray}

From Eqs.~$\left( \ref{coeff_A}\right) \ $and $\left( \ref{coeff_B}\right) ,$
we can set $\ (u^{\prime },v^{\prime },w^{\prime })\ =r^{\prime }(u,v,w)$,
because they satisfy the same equation. If both wave functions $\left( \ref
{wave_A}\right) $ and $\left( \ref{wave_B}\right) $ are true eigenfunctions
for a nondegenerate eigenstate with eigenvalue $E$, they should be
proportional with each other, so with a complex constant $r,\ $we can write
\begin{eqnarray}
\sum_{m=0}^\infty \sqrt{m!}u_m|m\rangle _A &=&r\sum_{m=0}^\infty (-1)^m\sqrt{%
m!}w_m|m\rangle _B, \\
\sum_{m=0}^\infty \sqrt{m!}v_m|m\rangle _A &=&r\sum_{m=0}^\infty (-1)^m\sqrt{%
m!}v_m|m\rangle _B, \\
\sum_{m=0}^\infty \sqrt{m!}w_m|m\rangle _A &=&r\sum_{m=0}^\infty (-1)^m\sqrt{%
m!}u_m|m\rangle _B.  \label{propotional}
\end{eqnarray}
Left multiplying $_A\langle n|$ followed by the use of $\ \sqrt{n!}%
\left\langle 0\right| |n\rangle _A=(-1)^n\sqrt{n!}\left\langle 0\right|
|n\rangle _B=e^{-g^2/2}g^n\ $ yields
\begin{equation}
\sum_{n=0}^\infty u_ng^n=r\sum_{n=0}^\infty w_ng^n,\ \sum_{n=0}^\infty
v_ng^n=r\sum_{n=0}^\infty v_ng^n,\ \sum_{n=0}^\infty
w_ng^n=r\sum_{n=0}^\infty u_ng^n.
\end{equation}
Eliminating the ratio constant $r$ gives $\ \left( \sum_{n=0}^\infty
u_ng^n\right) ^2=\left( \sum_{n=0}^\infty w_ng^n\right) ^2,\ $ then we
define a $G-$function as
\begin{equation}
G_{\pm }=\sum_{n=0}^\infty \left[ u_n\mp w_n\right] g^n,  \label{G_func}
\end{equation}
where $-\left( +\right) \ $ sign on the right hand side corresponds to even
(odd) parity''.

To this end, we can not define the $G$ function by one coefficient
determined through a linear three-term recurrence relation. To remedy this
problem, we will borrow the help of the third representation of the
wavefunction.

The wave function can be expanded in the Fock states, as in the one-qubit
QRM \cite{chen2012b} with a common feature of the conservation of parity
\begin{equation}
|\rangle =\left(
\begin{array}{c}
\sum_{n=0}^\infty \sqrt{n!}a_n|n\rangle  \\
\sum_{n=0}^\infty \sqrt{n!}b_n|n\rangle  \\
\pm \sum_{n=0}^\infty \sqrt{n!}\left( -1\right) ^na_n|n\rangle
\end{array}
\right) ,  \label{wave_d}
\end{equation}
where $+\left( -\right) $ stands for even (odd) parity. We then can get a
set of equations
\begin{eqnarray}
Ea_m &=&ma_m-\frac \Delta {\sqrt{2}}b_m+ga_{m-1}+g(m+1)a_{m+1}, \\
Eb_m &=&mb_m-\frac \Delta {\sqrt{2}}\left[ 1\pm \left( -1\right) ^m\right]
a_m,
\end{eqnarray}
which gives a linear three-term recurrence relation
\begin{equation}
a_{m+1}=\frac 1{g(m+1)}\left[ \left( E-m-\frac{\Delta ^2}{2\left( E-m\right)
}\left[ 1\pm \left( -1\right) ^m\right] \right) a_m-ga_{m-1}\right] .
\label{coeff_a}
\end{equation}
Setting $a_0=1$, we can obtain coefficients $a_m\ $ as a function of $E$
recursively, and $b_m$ are related to $a_m$ by
\begin{equation}
b_m=-\frac{\frac \Delta {\sqrt{2}}\left[ 1\pm \left( -1\right) ^m\right] }{%
E-m}a_m.  \label{coeff_b}
\end{equation}
Note that $b_{2k+1}=0$ for even parity, and $b_{2k}=0$ for odd parity.

Now, with the coefficients $a_m$, we can obtain all coefficients $u_m,v_m$
and $w_m$ with only one variable $E$. If both wave functions $\left( \ref
{wave_A}\right) $ and $\left( \ref{wave_d}\right) $ are true eigenfunctions
for a nondegenerate eigenstate with eigenvalue $E$, they should be in
principle only different by a complex constant $r^{\prime }$,
\begin{eqnarray}
\sum_{n=0}^\infty \sqrt{n!}u_n|n\rangle _A &=&r^{\prime }\sum_{n=0}^\infty
\sqrt{n!}a_n|n\rangle , \\
\sum_{n=0}^\infty \sqrt{n!}v_n|n\rangle _A &=&r^{\prime }\sum_{n=0}^\infty
\sqrt{n!}b_n|n\rangle , \\
\sum_{n=0}^\infty \sqrt{n!}w_n|n\rangle _A &=&\pm r^{\prime
}\sum_{n=0}^\infty \sqrt{n!}\left( -1\right) ^na_n|n\rangle .
\label{relation}
\end{eqnarray}
Projecting onto $_A\langle 0|$ yields the $n=0$ coefficients
\begin{eqnarray}
u_0 &=&\sum_{n=0}^\infty a_k\ \left( -g\right) ^k,  \label{coeff_u0} \\
v_0 &=&\sum_{k=0}^\infty b_{k\ }\left( -g\right) ^k,  \label{coeff_v0} \\
w_0 &=&\pm \sum_{k=0}^\infty \left( -1\right) ^ka_k\left( -g\right) ^k
\label{coeff_w0}
\end{eqnarray}
where use has been made of
\begin{equation}
_{\ A}\langle 0|k\rangle =\sqrt{\frac 1{k!}}e^{-g^2/2}\left( -g\right) ^k.
\end{equation}
Note that $r^{\prime }$ and $e^{-g^2/2}\ $are omitted if we are only
interested in the zeros of the G-function. Then for $n\geq 1$ coefficients $%
u_n(E),v_n(E)\ $and $w_n(E)\ $can be derived recursively
\begin{eqnarray}
v_n &=&-\frac 1{gn}\left[ \left( E-g^2-n+1\right) v_{n-1}+\frac \Delta {%
\sqrt{2}}(u_{n-1}+w_{n-1})+gv_{n-2}\right] ,  \label{coeff_v} \\
w_n &=&-\frac 1{2gn}\left[ (E-3g^2-n+1)w_{n-1}+\frac \Delta {\sqrt{2}%
}v_{n-1}+2gw_{n-2}\right] ,  \label{coeff_w} \\
u_n &=&-\frac{\frac \Delta {\sqrt{2}}}{E-n+g^2}v_n.  \label{coeff_u}
\end{eqnarray}
Inserting these coefficients to Eq.~(\ref{G_func}) finally yields a very
concise G function $G_{\pm }(E)$ without the use of a high dimensional
matrix. For any $E$, \ the coefficients $u_n(E),\ v_n(E)$, and $w_n(E)$ can
be expressed by only $a_k$ which is determined   in Eq.~( \ref{coeff_a}). In
this sense, this $G$ function is a transcendental function well defined
mathematically in a controlled way. The solutions to the model can be given
based on these simple $G$ functions with only one variable.

\section{ Solutions and discussions}

\begin{figure}[tbp]
\includegraphics[width=10cm]{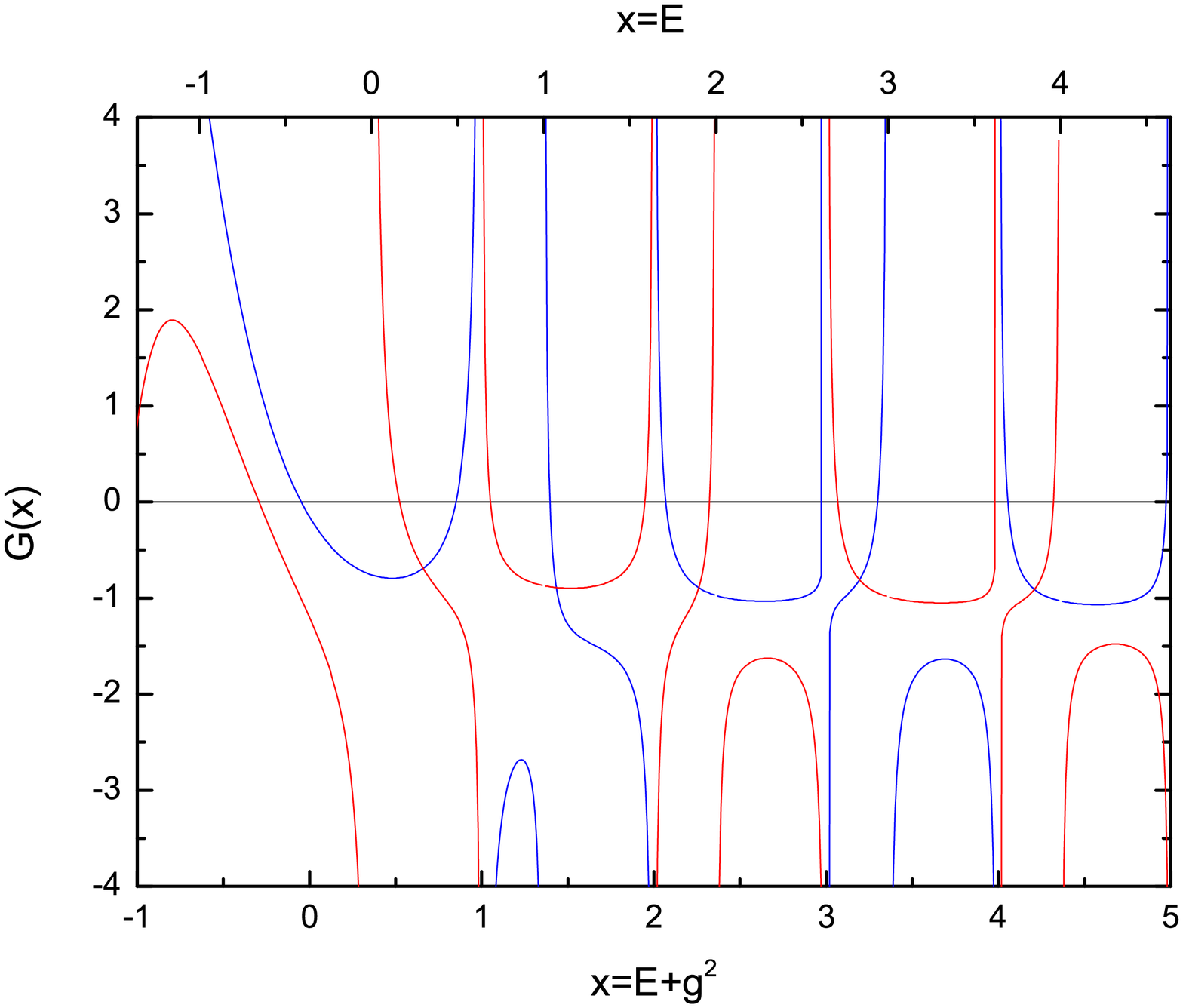}
\caption{(Color online) G(x) curves for the two-qubit QRM at $\Delta
=0.5,g=0.6$. The red (blue) one denotes even (odd) parity. }
\label{gfunction}
\end{figure}

\begin{figure}[tbp]
\includegraphics[width=10cm]{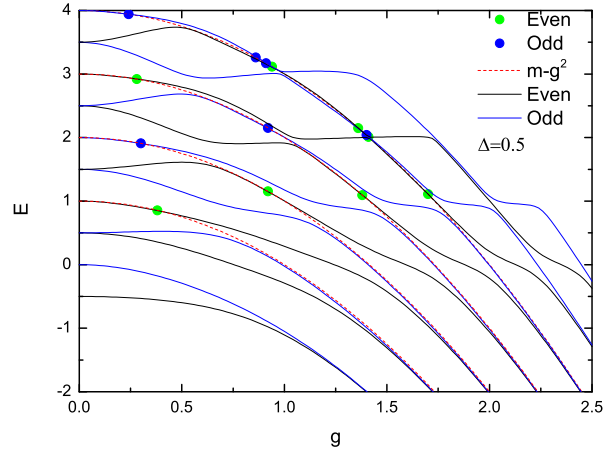}
\caption{(Color online) The spectrum and the isolated exceptional solutions
for the two-qubit QRM at $\Delta =0.5$. }
\label{spectrum05}
\end{figure}

First we plot the G function in Fig.~\ref{gfunction} for the case of $\Delta
=0.5$ and $g=0.6$. As the zeros of the G function yield the eigenenergies,
the exact energy levels are thus obtained as a function of the coupling
constant $g$, which are then displayed in Fig.~\ref{spectrum05} for qubit
splitting $\Delta=0.5$.

From Eqs.~(\ref{G_func}), (\ref{coeff_b}), (\ref{coeff_v0}), and (\ref
{coeff_u}), it is obvious that the G function is not analytic at two types
of poles, $E=n-g^{2}$ and $E=n$, a fact that is also clearly demonstrated in
Fig.~\ref{gfunction}. The first type of poles are precisely the eigenvalues
of the uncoupled bosonic mode of vanishing qubit splitting ($\Delta=0$),
similar to those in one-qubit counterpart\cite{Braak}. It follows that
isolated exact solutions may also exist. The second type of poles are the
trivial eigenvalue in the subspace $\left\vert j=0,m_z=0\right\rangle $.
Such poles are, however, absent in the one-qubit QRM.

In the one-qubit QRM, Koc \textit{et al.}~have obtained isolated exact
solutions\cite{Koc}, which are the Juddian solutions \cite{Judd} with doubly
degenerate eigenvalues. These analytical solution at some isolated points
can serve as benchmarks to test various approximate approaches developed for
the entire parameter space. In the context of the $G$ function, these
isolated eigenvalues do not correspond to the zeros of the $G$ functions,
and are therefore called exceptional solutions\cite{Braak}. In the
two-photon QRM, these isolated exact solutions\cite{Bishop} are elusive in
the absence of the single-variable G functions\cite{Trav}. It is however
easily derived later on the basis of the single variable G function\cite
{Chen2012}. In the scheme of ECS, we know that these isolated solutions with
degenerate eigenstates are naturally excluded from the zeros of $G $
function on the basis of proportionality. In this work, the single variable
G-function obtained for the two-qubit QRM allows an in-depth discussion of
the two types of poles.

(1) \textsl{Exceptional solutions at $E=m-g^2$}. On the first type of poles
at $E=m-g^2$, the necessary and sufficient condition for the occurrence of
the exceptional eigenvalue is that the numerator ($v_m)\ $in $u_m$ given by
Eq.~(\ref{coeff_u}) vanishes so that the pole at $x=m$ is lifted. Eq.~(\ref
{relation}) gives the condition for which the poles are removed:
\begin{equation}
\sum_{n=0}^\infty \sqrt{n!}\ b_n^{(\pm )}D_{mn}=0,  \label{exceptional}
\end{equation}
where
\begin{equation}
D_{mn}=\left( -g\right) ^{n-m}\sqrt{\frac{m!}{n!}}L_m^{n-m}(g^2)=\left(
-1\right) ^{n-m}D_{nm}.
\end{equation}
Here, for $m\leq n,\ L_m^{n-m}(g^2)$ is Laguerre polynomial,$\ b_n^{(\pm )}\
$ is determined through Eqs.~(\ref{coeff_a}) and (\ref{coeff_b}) with even
and odd parity. For the same $\Delta $, the numerator $v_m$ is parity
dependent, therefore in general, $g_m^{(+)}\neq g_m^{(-)}$, in sharp
contrast with one-qubit QRM with both one photon \cite{Braak} and two
photons \cite{Chen2012}, where the numerator is parity independent.

Using the condition (\ref{exceptional}) of the exceptional solution, one can
obtain $g_{m}^{(\pm )}$ at $E=m-g^{2}$ without the help of the $G$ function.
The positions of the isolated solutions are also collected in Fig.~2. Very
interestingly, they are just the crossing points of curves for $E=m-g^{2}$
and the corresponding energy levels $E^{\pm }(g)$, indicating that at least
one type of exceptional solutions in this model has the form $E=m-g^{2}$.

To show the characteristics of the G function at these isolated points, in
Fig.~3 we plot $G(x)$ with $E=2-g^{2}$ at given $g_{m=2}^{(\pm)}$ for the
case of $\Delta =0.5$. It is clearly demonstrated that the pole at $x=2$ is
removed due to the vanishing numerator, and $x=2$ is an exceptional solution
because $G_{\pm }(x=2)\neq 0$. However, the poles of $G_{-}(x)$ at $%
g_{m=2}^{(+)}$ and of $G_{+}(x)$ at $g_{m=2}^{(-)}$ remain. These
exceptional solutions are not the degenerate eigenvalues, at variance with
those in the one-qubit QRM. Without degeneracy, $G_{+}(x)$ must be zero at $%
g_{m}^{(+)}$ if the eigenstate is given by Eq.~(\ref{wave_A}). It is
concluded that the wave function can not be described by an expansion of the
displaced operator $A=d+g$ as in Eq.~(\ref{wave_A}) at these isolated points
shown as filled circles in Fig.~\ref{spectrum05}.

\begin{figure}[tbp]
\includegraphics[width=6cm]{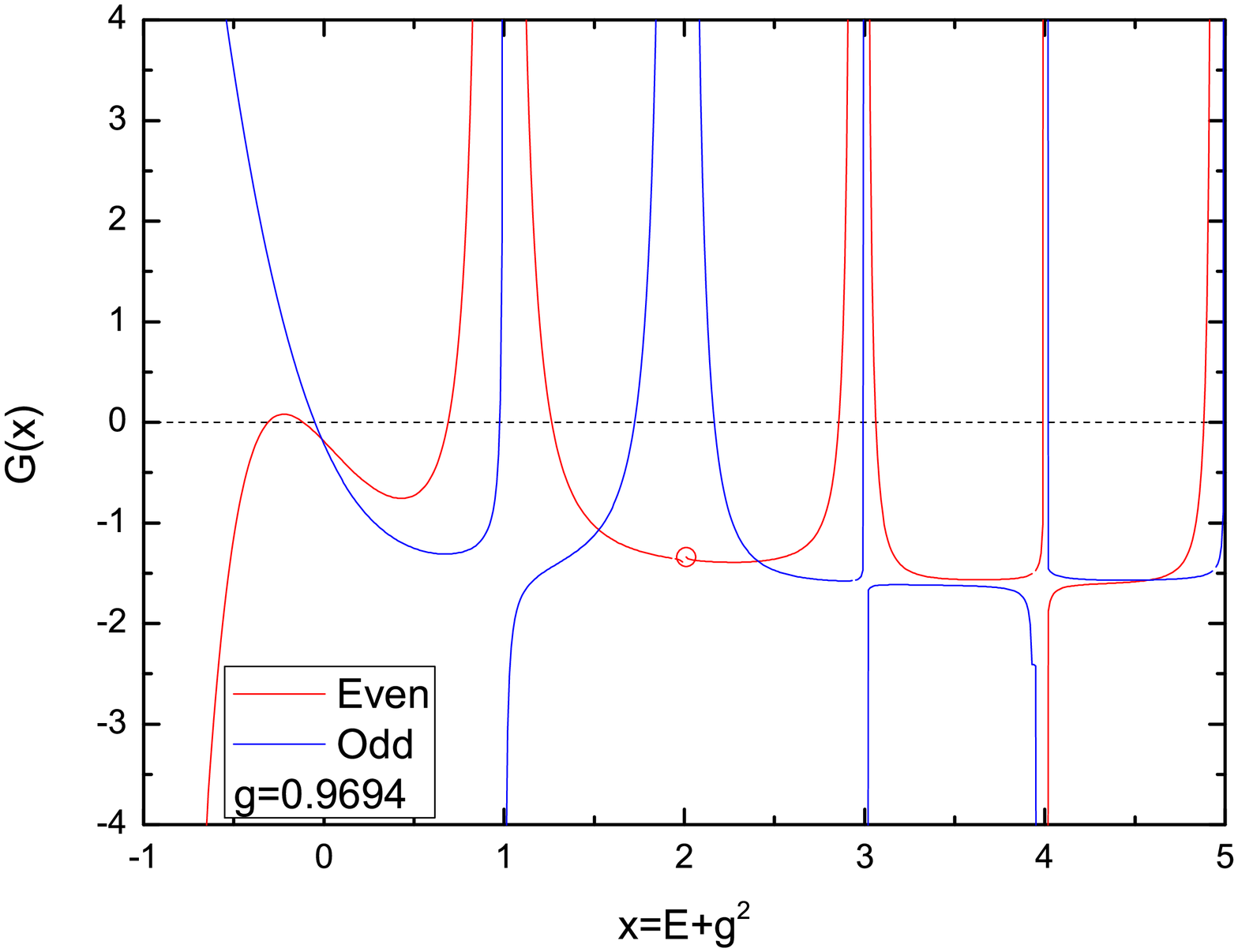} %
\includegraphics[width=6cm]{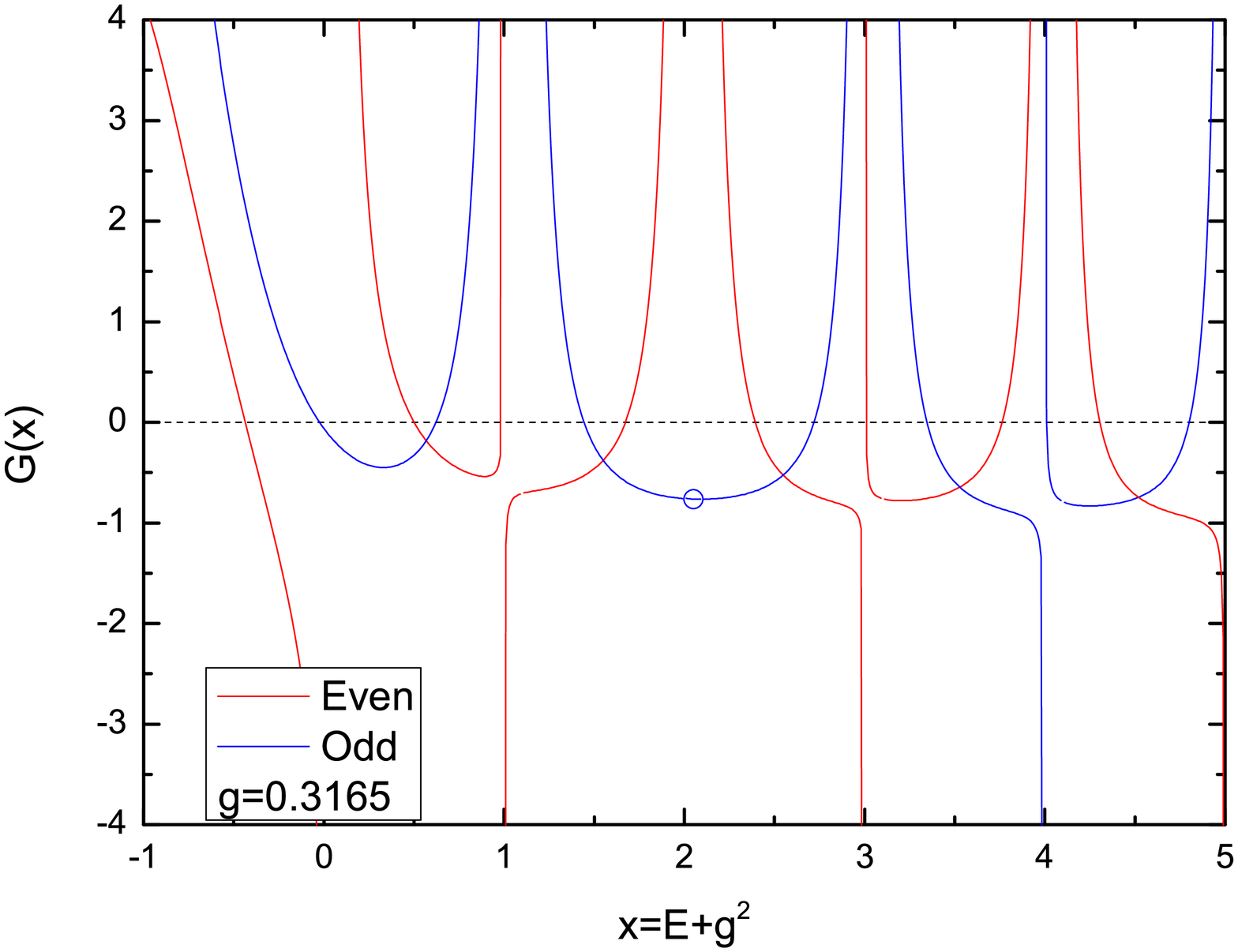}
\caption{(Color online) G(x) at exceptional solutions for $E=2-g^{2}$. Other
parameters are the same as in Fig.~2. }
\label{except}
\end{figure}

In the one-qubit QRM, the exceptional eigenvalues occur at some level
crossing points\cite{Braak,Chen2012}, the proportionality between the two
wave functions in these doubly degenerate states is lacking, we may not have
$G_{\pm }(x)=0$, and so the doubly degenerate eigenstate is still
describable in terms of operator $A$ with displacements $\pm g$ as in Eqs.~(
\ref{wave_A}) and ~(\ref{wave_B}). For the two-qubit QRM, however, this is
not the case. In other words, the exact solutions exist for the one-qubit
QRM, in contrast to quasi-exact solutions in the two-qubit QRM. This
difference may be responsible for the integrability in one-qubit QRM and
quasi-integrability in the Dicke model with $N\ge 2$.

(2) \textsl{Isolated poles at $E=n$}. Is there a second kind of exceptional
solutions of the form $E=n$? In the recursive relation Eqs.~(\ref{coeff_a})
and (\ref{coeff_b}), we note that the coefficient diverges for even $n$ in
the even parity and odd $n$ in the odd parity, consistent with the behavior
of the calculated $G$ function in Fig.~\ref{gfunction}. Can the pole at $E=n$
be removed for a certain model parameter so that the G function at this
point remains analytical? It is required that numerator $a_n$ in Eq.~(\ref
{coeff_b}) vanishes at $E=n$. For $n=0,$ $b_0=-\sqrt{2}\Delta /E$ diverges
at $E=0$ for even parity, while for $n=1,$ $b_1=\frac{\sqrt{2}E\Delta }{%
g\left( E-1\right) }$ diverges at $E=1$ for odd parity. So at $E=0,1$, $G(x)$
always diverges for any model parameters, and there exists no exceptional
solution. For $n=2,\ b_n$ diverges at $E=2$ for even parity as
\begin{equation}
b_2=-\frac \Delta {\sqrt{2}g^2}\frac{\left( E-1\right) \left( E-\frac{\Delta
^2}E\right) +g^2}{\left( E-2\right) }.
\end{equation}
Hence it is only possible for the numerator to vanish at $g_2=\sqrt{\frac{%
\Delta ^2-4}2}$ for $\Delta \geq 2$. But it is found in our calculations
that not even $g_2$ is at the crossing point of the $E=2$ line and the
energy levels. Therefore, $E=n$ is only an isolated pole, not an exceptional
solution, in sharp contrast to  $E=n-g^2$. It makes sense physically. The
eigenvalue $E=n$ of the spin singlet state $\left| n\right\rangle =\left|
0,0\right\rangle \left| n\right\rangle _{\mathrm{ph}}$ should be independent
of the nontrivial states in the $j=1$ subspace. So it is impossible that $E=n
$ is an exceptional solution for the G functions derived in the $j=1$
subspace. It only points to some intrinsic relations between the two
subspaces sharing the common quantum number $m_z=0$ in the the same system.
In addition, it is our belief that these poles are not the origin of the
quasi-integrability of the QRM with two qubits, because they are also
well-defined eigenvalues of this model.

\section{Conclusions}

In this work, we have derived for the QRM with two equivalent qubits a
concise single-variable G function which leads to simple, analytical
solutions. Two type of poles in the G function are observed. One is the
eigenvalue of the uncoupled bosonic mode, also known as the exceptional
solution. Obtained analytically is the condition for the occurrence of the
exceptional solution, which is parity dependent, in sharp contrast to its
one-qubit counterpart, the Juddian solution to the one-qubit QRM. The other
type of poles occurs at the eigenvalue of the spin singlet state, which is
not an exceptional solution. This work extends the methodology of a compact
G function in the one-qubit QRM to the two-qubit case, thereby allowing a
conceptually clear, practically feasible treatment to energy spectra. It is
our expectation that the present approach will find more applications in the
future, such as in real-time dynamics.

\acknowledgments This work was supported by National Natural Science
Foundation of China under Grant No. 11174254, National Basic Research
Program of China (Grant Nos. 2011CBA00103 and 2009CB929104), and the
Singapore National Research Foundation through the Competitive Research
Programme (CRP) under Project No. NRF-CRP5-2009-04.

\end{document}